\documentclass{article}
\usepackage{spconf,amsmath,graphicx}
\usepackage{subcaption}
\usepackage[T1]{fontenc}

\graphicspath{{Figs/}}

\title{EXPLORING HERITABILITY OF FUNCTIONAL BRAIN NETWORKS WITH INEXACT GRAPH MATCHING}
\name{Sofia Ira Ktena, Salim Arslan, Sarah Parisot, Daniel Rueckert}
\address{Biomedical Image Analysis Group, Imperial College London, UK}
\begin{document}
%
\maketitle
\begin{abstract}
Data-driven brain parcellations aim to provide a more accurate representation of an individual's functional connectivity, since they are able to capture individual variability that arises due to development or disease. This renders comparisons between the emerging brain connectivity networks more challenging, since correspondences between their elements are not preserved. Unveiling these correspondences is of major importance to keep track of local functional connectivity changes. We propose a novel method based on graph edit distance for the comparison of brain graphs directly in their domain, that can accurately reflect similarities between individual networks while providing the network element correspondences. This method is validated on a dataset of 116 twin subjects provided by the Human Connectome Project.
\end{abstract}
\begin{keywords}
functional brain connectivity, twin study, graph matching
\end{keywords}
\section{Introduction}
\label{sec:intro}

The extremely complex circuit of our central nervous system constitutes the primary means of information transmission within the brain, and is responsible for different cognitive functions. 
Structural brain connectivity can be examined at the macroscale to obtain a complete map of the neuroanatomical connections  between brain regions
, the so-called human connectome~\cite{sporns2005human}. Functional connectivity, in turn, can be used to explore the temporal dependency between neurophysiological events and expresses network behaviour underlying high level cognitive processes~\cite{biswal1995functional}.

Individual differences in network topology of the connectome have previously been associated with heritable phenotypes such as intelligence~\cite{li2009brain}
, giving rise to questions regarding the heritability of network topology itself. Addressing these questions would be a leap forward in understanding mechanisms through which genetic influences on brain morphology eventually contribute to human behaviour in health and disease. In this context, twin designs have recently been used to study the genetic influence on both structural~\cite{bohlken2014heritability, zhan2015heritability} and functional~\cite{
van2013genetic, fu2015genetic, yang2016genetic} brain connectivity. More specifically, Bohlken et al.~\cite{bohlken2014heritability} studied the extent to which genes shape the topology of structural brain connectivity. They estimated a significant heritability index for normalised clustering coefficient and characteristic path length. A similar study~\cite{zhan2015heritability} performed probabilistic tractography on healthy adults and obtained similarly significant heritability for global efficiency and network strength. Global efficiency of functional networks was also found to be under genetic control in~\cite{van2013genetic}, while Fu et al.~\cite{fu2015genetic} tried to gain insight into the genetic basis of resting state networks (RSNs) with a voxelwise genetic analysis. Yang et al.~\cite{yang2016genetic} explored the heritability of intrinsic connectivity networks (ICNs) and found interactions between them to be mostly influenced by environmental factors, despite the fact that the activity of most ICNs is driven by genetic factors.

In all studies mentioned above, brain parcellations have been employed to construct lower dimensional graph representations of the brain networks. 
The delineation of distinct brain regions is often based on information derived from anatomical atlases or regions of interest reported in the literature~\cite{salvador2005neurophysiological
}. However, anatomical landmarks like the sulci and gyri might not necessarily align with functional boundaries between brain regions, hence data-driven approaches have also been used to parcellate the brain. These can be group-based, ensuring correspondences between subjects
, or tailored for individual subjects to account for inter-subject variability and discard the constraints of standardised anatomy~\cite{
arslan2015joint, parisot2016grampa}. In the latter case, however, correspondences across parcellations/subjects are not preserved, while individual networks might even entail a different number of nodes. Thus, these methods require more sophisticated approaches for graph comparison than the standard network measures used in the previous studies.

This problem is addressed in~\cite{tymofiyeva2014brain} by aligning the individual connectivity matrices with simulated annealing to study the differences in structural connectivity between age groups. However, this method does not account for any characteristic information of a region, other than its connectivity profile to the rest of the brain. A graph kernel, which captures structural, geometric and functional information, is used in~\cite{takerkart2014graph} to evaluate similarity between graphs. Nevertheless, the latter does not provide element correspondences of the networks compared. In this work we propose a method based on graph edit distance to assess brain graph similarity and obtain element correspondences between individual subject networks and show that it can successfully reflect genetic similarities.

\section{METHODS}
\label{sec:methods}

\subsection{Dataset and preprocessing}
The dataset used in this study is minimally preprocessed and provided by the Human Connectome Project~\cite{glasser2013minimal}. It consists of 50 healthy unrelated subjects, 66 female monozygotic (MZ) twins and 50 female dizygotic (DZ) twins. The available data for each subject include two resting-state fMRI (rs-fMRI) sessions that are preprocessed to remove spatial artifacts and distortions and are, finally, converted to a standard ``grayordinate'' space to facilitate cross-subject comparisons. Additionally, the time series for each grayordinate is denoised and normalised to zero mean and unit variance.

\begin{figure}[t]

\begin{minipage}[b]{.21\linewidth}
  \centering
 \centerline{\includegraphics[width=\linewidth]{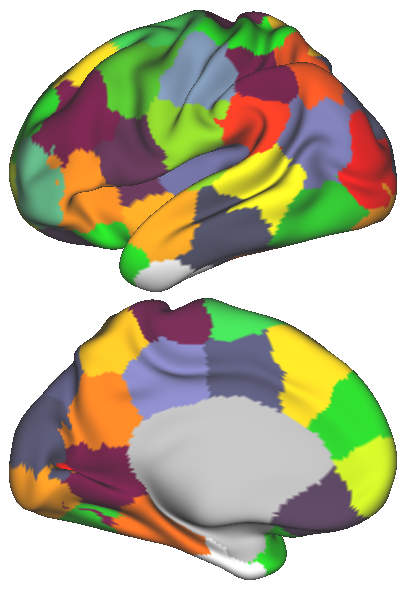}}
\end{minipage}
\hspace{0.2cm}
\begin{minipage}[b]{.21\linewidth}
  \centering
 \centerline{\includegraphics[width=\linewidth]{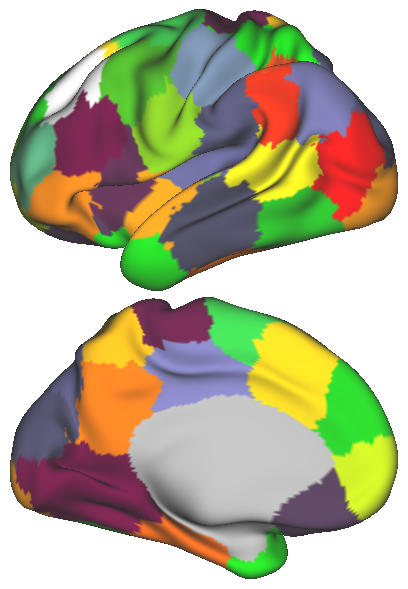}}
\end{minipage}
\hspace{0.2cm}
\begin{minipage}[b]{0.21\linewidth}
  \centering
 \centerline{\includegraphics[width=\linewidth]{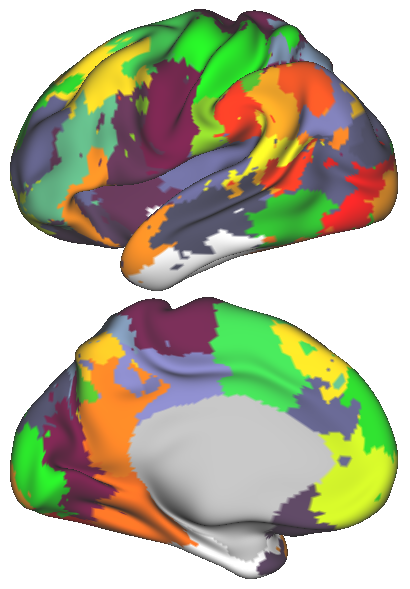}}
\end{minipage}
\hspace{0.2cm}
\begin{minipage}[b]{0.21\linewidth}
  \centering
 \centerline{\includegraphics[width=\linewidth]{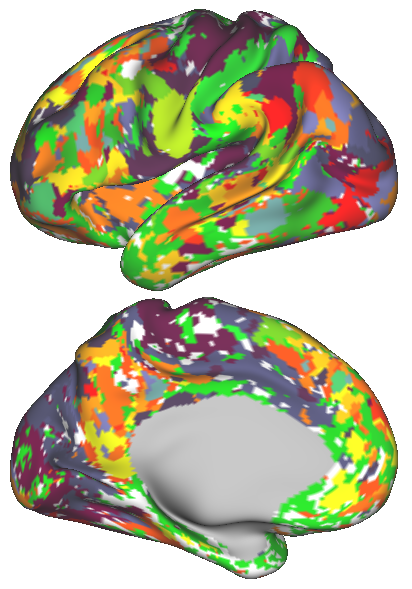}}
\end{minipage}
\caption{Parcellations of an individual's left hemisphere obtained with k-means clustering (the contribution of rs-fMRI timeseries against spatial coordinates is $c=10, 50, 75, 100\%$ from left to right, while the number of parcels remains the same). The higher the contribution of rs-fMRI time series the less the smoothness and spatial contiguity of the parcels, i.e. parcels may be spatially scattered across the hemisphere.}
\label{fig:parc}
\end{figure}

\subsection{Network construction}
In order to obtain a network representation of an individual's functional brain connectivity, the network elements, i.e. nodes and edges, need to be defined. In this study, k-means clustering is applied to the rs-fMRI data~\cite{thirion2014fmri}, in order to parcellate the brain into $N$ regions with distinct rs-fMRI activation patterns. The number of parcels, $N$, may vary and determines the number of nodes within the network. The spatial coordinates can also be taken into account along with the fMRI time series during clustering. The effect of introducing such spatial information on the parcellation is illustrated in Fig.~\ref{fig:parc}. Subsequently, the average time series within each parcel is used to represent the corresponding node. The connection strength between two nodes, namely the edge weight, is estimated from the partial correlation of the representative time series, in order to discard the indirect functional connections between two cortical regions.
 
\subsection{Similarity estimates}
\textbf{\textit{1) Matrix alignment with simulated annealing (SA).}}
Simulated annealing was used in~\cite{tymofiyeva2014brain} to align structural connectivity networks generated with random parcellations. It is a commonly used method for multiple sequence alignment of biological sequences. In this approach, a series of node reorderings is attempted in one of the adjacency matrices, in order to find a better alignment that minimizes the distance metric between the two matrices, which in this case is the absolute or square distance. Simulated annealing is a global optimization technique which uses a  ``temperature factor'' that determines the rate at which reorderings take place and the likelihood of each reordering. The implementation is based on the matrix alignment function of the Brain Connectivity Toolbox (http://www.brain-connectivity-toolbox.net~\cite{rubinov2010complex}).

\begin{figure}[t]
\centering
\includegraphics[width=\linewidth]{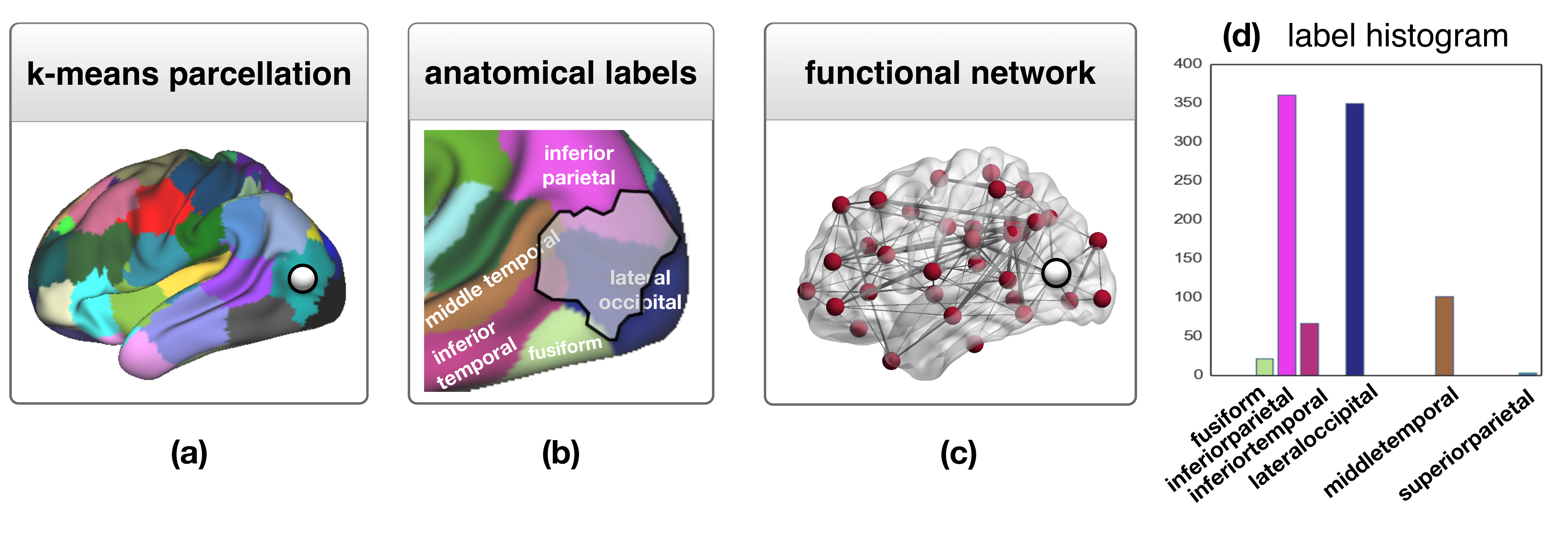}
\vspace{-0.7cm}
\caption{\textbf{(a)} Single-subject parcellation of an individual's cortex with a parcel $p_i$ covering part of the occipital, temporal and parietal lobes highlighted. \textbf{(b)} Anatomical labels from the Desikan atlas for the same subject, with the boundary of the region corresponding to $p_i$ outlined. \textbf{(c)} Functional network with nodes obtained using parcellation \textit{(a)} and edge weights estimated with partial correlation between the representative rs-fMRI time series of the connected nodes; the node $n_i$ corresponding to $p_i$ is highlighted with a gray color. \textbf{(d)} Anatomical label histogram for node $n_i$.}
\label{fig:labels}
\end{figure}

\textbf{\textit{2) Graph edit distance (GED)}}. Graph edit distance is a non-negative function that evaluates the similarity between two graphs directly in their domain $\mathcal{G}$ and is able to model structural network variations in an intuitive way. In order to calculate the graph edit distance, the minimum cost edit path that transforms one graph into another needs to be estimated. The total cost of this edit path is defined as the sum of all edit operations, which in this case might include node and edge additions, deletions and substitutions:
 
\begin{equation}
d_{GED}(G, G') = \min_{o \in \mathcal{O}} \sum_i c(o_i)
\end{equation}

where $c(o_i)$ is the cost of edit operation $o_i$,  $o = (o_1, \ldots, o_k)$ is an edit sequence from $G$ to $G'$, and $\mathcal{O}$ the finite set of edit sequences from $G$ to $G'$.

Graph edit distance allows us to incorporate label information about the network elements in the distance estimate. In the proposed approach, the histogram of anatomical labels of the voxels constituting a parcel is used to characterize each node (Fig.~\ref{fig:labels}).

\textit{Edit costs.} The costs of edit operations are defined as follows: a cost of $|e_{ij}|$ is used for inserting or deleting an edge
 of weight $e_{ij}$. The cost of node insertion or deletion is equal to the cost of insertion or deletion of all edges attached to it, i.e. $c_{i \to \varepsilon} = c_ {\varepsilon \to i} = \sum_j |e_{ij}|$. Moreover, the cost of modifying edge $e_{ij}$ to $e'_{ij}$ (i.e. edge substitution), is weighted by the cosine distance $d_{cosine}$ of the corresponding node label histograms. This means that if node $i$ of $G$ with label histogram $h_i=(l_{i1}, \ldots, l_{im})$ is substituted by node $i'$ of $G'$ with label histogram $h_{i'}=(l_{i'1}, \ldots, l_{i'm})$, then $c_{e_{ij} \to e'_{ij}} = d_{cosine}(h_i, h_{i'})\times|e_{ij}-e'_{ij}|$ for all $j, j'$. Node substitutions are only allowed between nodes on the same hemisphere and with $d_{cosine}(h_i, h_{i'}) < 1$, hence the cost of any other substitution is infinite.

It should be noted that the exact GED computation is exponential in terms of time and space to the number of nodes involved, making its application computationally prohibitive when the dimensionality of the network increases. Therefore, an approximate estimate of GED is used instead, which is based on minimum cost bipartite matching~\cite{riesen2007bipartite}.

\section{RESULTS}
\label{sec:results}

\subsection{Estimated distances}
\begin{figure}[t]
\centering
    \begin{subfigure}[b]{0.49\linewidth}
        \includegraphics[width=\textwidth]{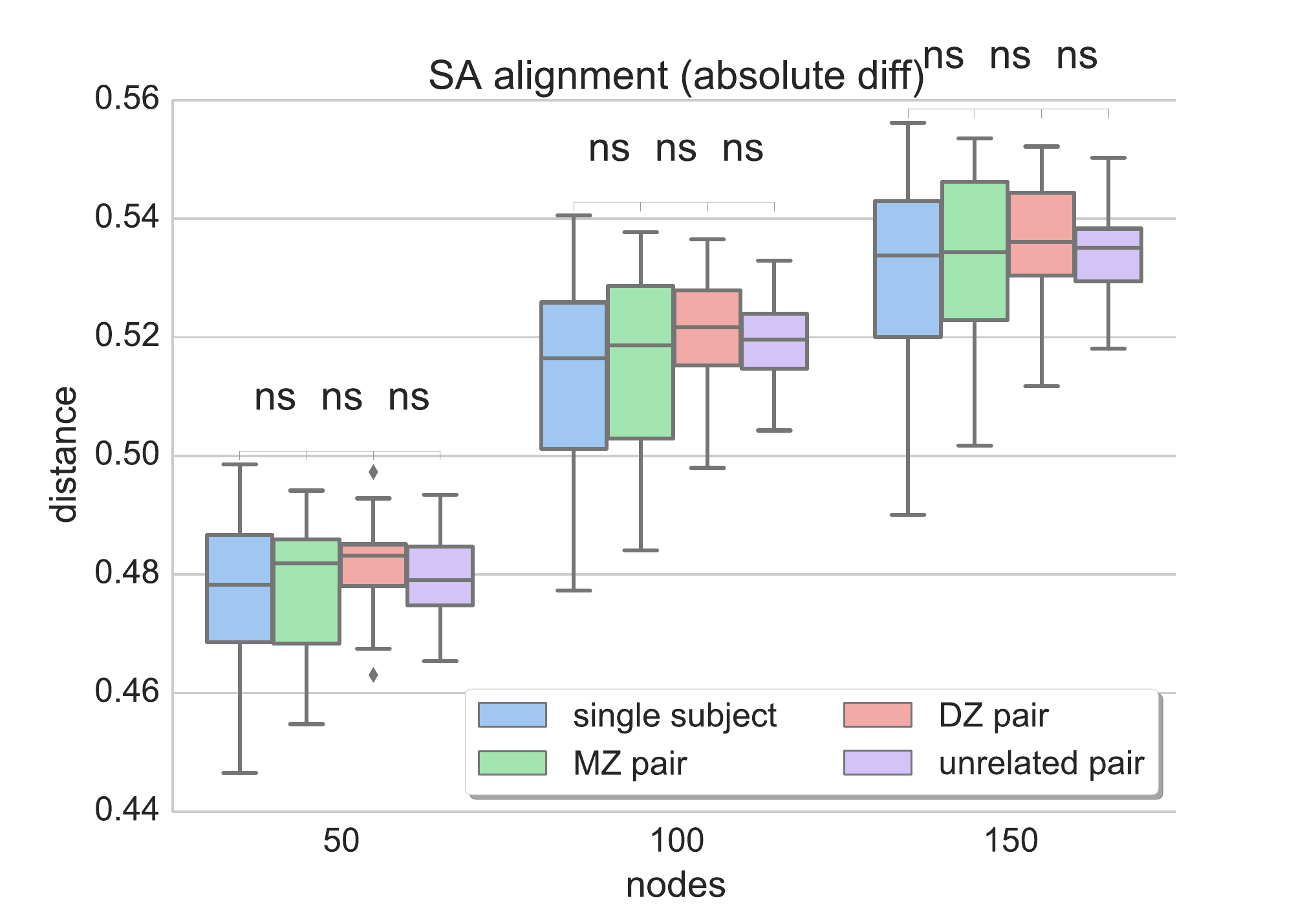}
	\end{subfigure}
    \begin{subfigure}[b]{0.49\linewidth}
        \includegraphics[width=\textwidth]{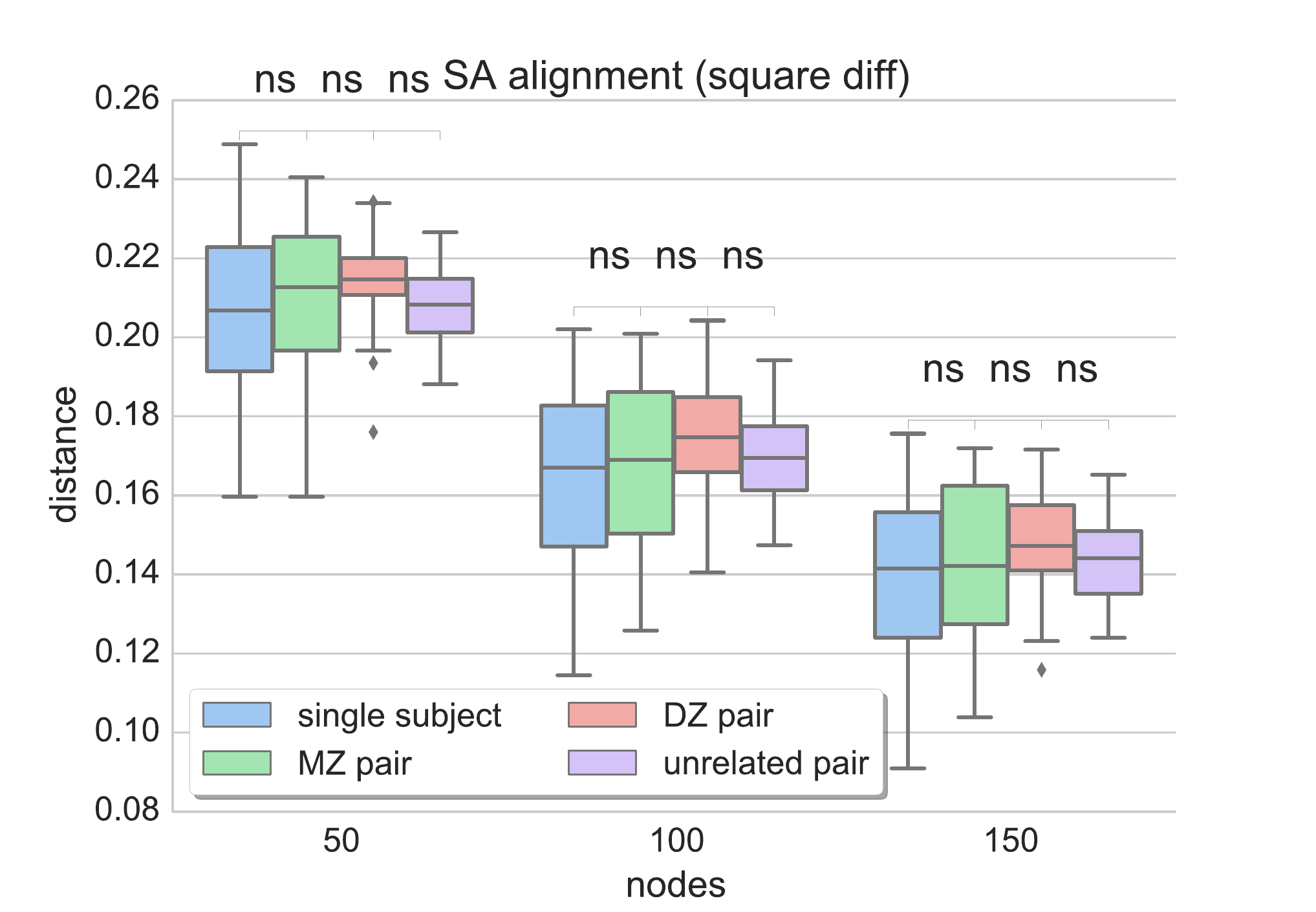}
	\end{subfigure}
\caption{Minimum distances obtained with SA on networks derived from parcellations using only fMRI data ($c=100\%$); absolute (left) and square difference (right) were used to estimate distance between two connectivity matrices. Results are similar for parcellations with different fMRI contributions.}
\label{fig:boxplots_ma}
\end{figure}

\begin{figure*}[t]
\centering
\includegraphics[width=0.83\textwidth]{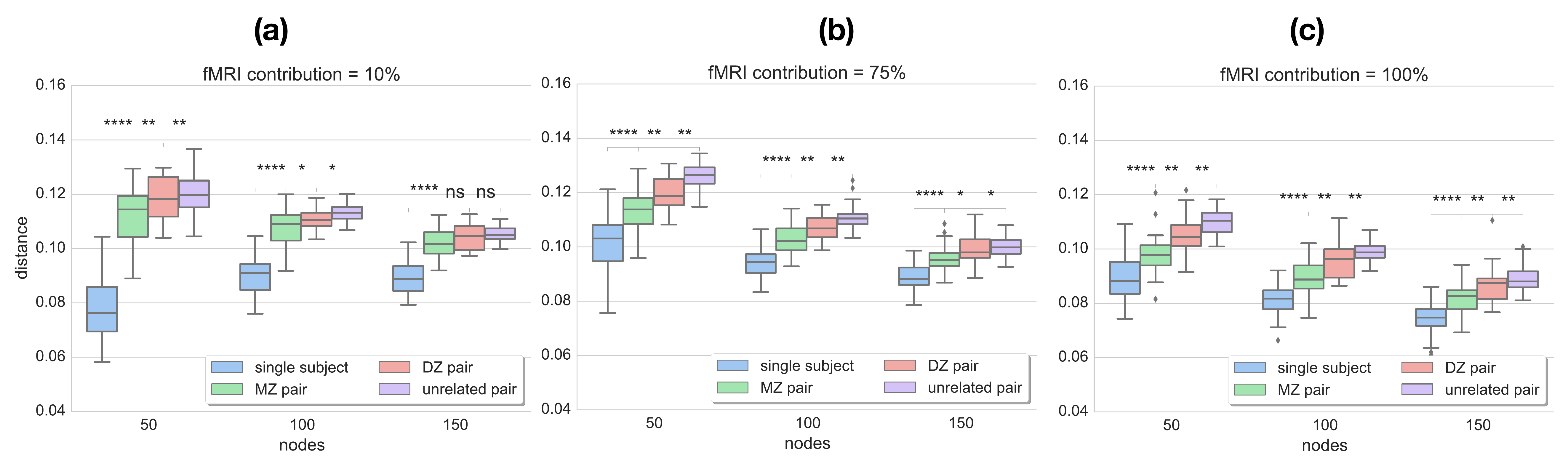}
\caption{GED values calculated for networks of size $N=50, 100, 150$ and varying contribution of rs-fMRI data to the parcellation. Permutation test results are shown as non-significant (n.s.), $p<0.05$ (*), $p<0.01$ (**), $p<0.0001$ (****).}
\label{fig:boxplots_ged}
\end{figure*}

Since the provided dataset consists of two rs-fMRI sessions (acquired 1 day apart), two different data-driven parcellations and functional connectivity networks can be generated for each individual. The distances between two networks of the same subject are compared to the distances estimated between all MZ and DZ twin pairs, as well as unrelated pairs. Ideally, the distance measure should be able to reflect the similarities due to genetic factors, i.e. yield lower distances for MZ than DZ twin pairs, while distances between networks of the same subject should be even lower. We compare the results obtained with matrix alignment using simulated annealing and the proposed method (graph edit distance) along different levels of parcellation granularity and fMRI data contribution (for a direct comparison see Figs.~\ref{fig:boxplots_ma} and~\ref{fig:boxplots_ged}c).

Since the distance estimate as well as the final alignment of the matrices in the SA approach is dependent on the initial alignment of the matrices, 100 random permutations of the connectivity matrix are computed and fed to the matrix alignment algorithm. In the end, the alignment that yields the minimum cost across the different runs is preserved and plotted in Fig.~\ref{fig:boxplots_ma}. It can be observed that, although the median distance tends to be higher between MZ pairs compared to single subject networks as well as between DZ pairs compared to MZ, the mean values obtained with SA do not provide a significant separation between these populations. Also, distances between unrelated pairs are almost at the same level as MZ pairs. This result is consistent across different resolutions, regardless of the contiguity of the parcels.

On the contrary, GED estimates capture the greater similarity that is expected between networks of the same subject, as well as between MZ pairs compared to DZ and unrelated pairs (Fig.~\ref{fig:boxplots_ged}). It can also be observed that the more data-driven the parcellation is (i.e. less effect of spatial coordinates), the better the separation between MZ and DZ twin pairs. This indicates that the parcellation describes individual connectivity more accurately, while the specific distance measure successfully reflects that.

\subsection{Network element correspondences}
\begin{figure}[t]
\centering
\includegraphics[width=.72\linewidth]{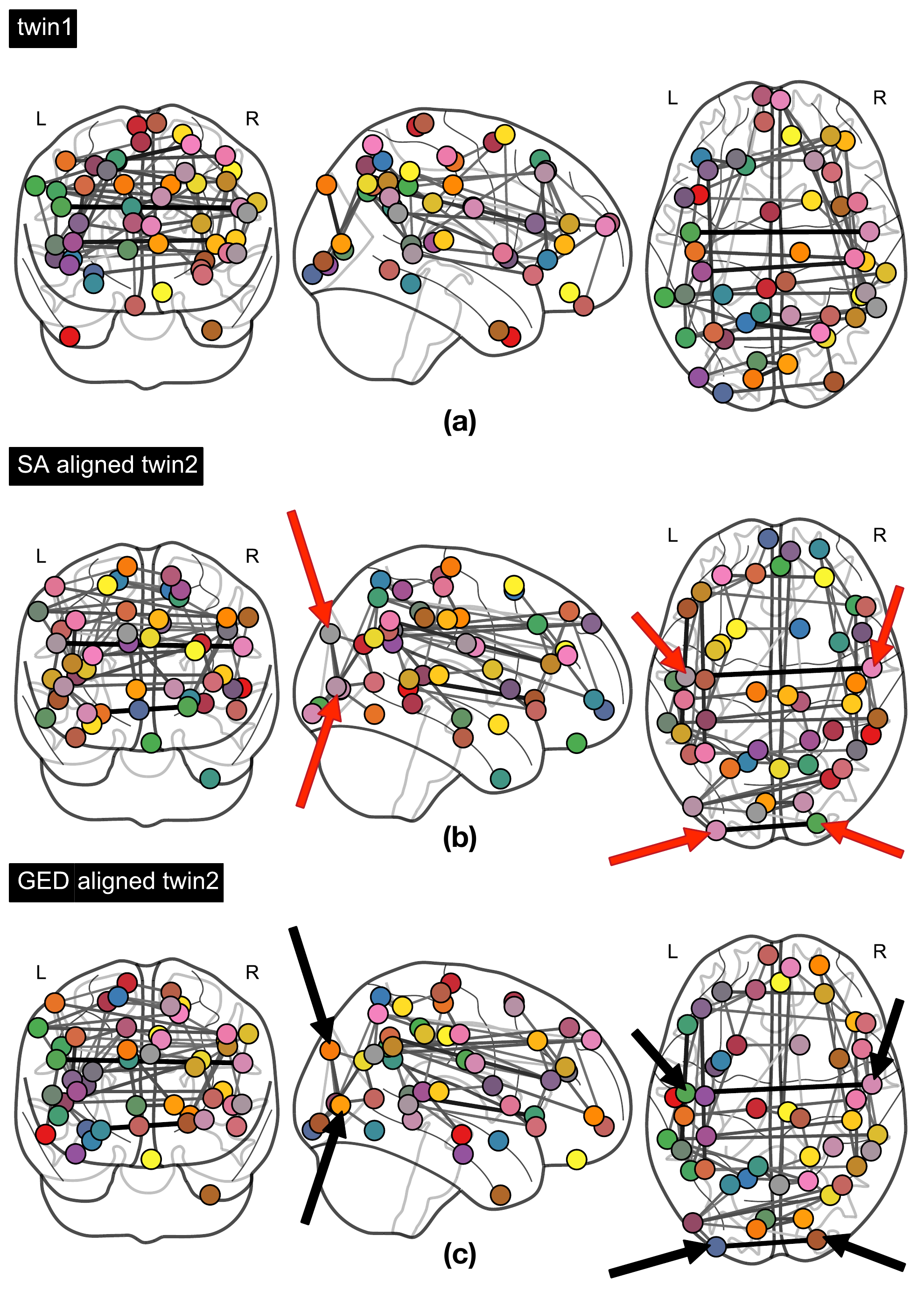}
\caption{Node correspondences indicated with matching colors between twin1 \textbf{(a)} and twin2 for a MZ pair obtained with matrix alignment \textbf{(b)} and graph edit distance \textbf{(c)}. Red arrows show mismatched nodes/node swaps from different hemispheres. Black arrows show correctly matched nodes.}
\label{fig:correspondences}
\end{figure}

The node correspondences obtained with the two different distance measures for a MZ twin pair are visually compared in Fig.~\ref{fig:correspondences}. It can be observed that since SA is unaware of the node labels, swaps between nodes from different hemispheres can take place (red arrows). Nevertheless, in the case of GED correspondences, nodes are matched accurately, with important functional connections maintained between matched nodes for both twin subjects (black arrows).

\section{CONCLUSION}
\label{sec:conclusion}
We propose a novel approach for evaluating similarities between functional brain networks, derived from individual subject parcellations. The method successfully reflects greater similarities between networks from the same subject, as well as from MZ twin pairs compared to DZ and unrelated pairs. It also correctly identifies correspondences between network elements paving the way for comparisons between healthy and diseased populations using data-driven individual networks that might even consist of different numbers of nodes.



\bibliographystyle{IEEEbib}
\renewcommand{\refname}{}
\bibliography{refs}

\begin{thebibliography}{10}

\bibitem{sporns2005human}
O~Sporns, G~Tononi, and R~K{\"o}tter,
\newblock ``The human connectome: a structural description of the human
  brain,''
\newblock {\em PLoS Comput Biol}, vol. 1, no. 4, pp. e42, 2005.

\bibitem{biswal1995functional}
B~Biswal et~al.,
\newblock ``Functional connectivity in the motor cortex of resting human brain
  using echo-planar {MRI},''
\newblock {\em Magn Res in Medicine}, vol. 34, no. 4, pp. 537--541, 1995.

\bibitem{li2009brain}
Y~Li et~al.,
\newblock ``Brain anatomical network and intelligence,''
\newblock {\em PLoS Comput Biol}, vol. 5, no. 5, pp. e1000395, 2009.

\bibitem{bohlken2014heritability}
MM~Bohlken et~al.,
\newblock ``Heritability of structural brain network topology: a {DTI} study of
  156 twins,''
\newblock {\em Hum Brain Mapp}, vol. 35, no. 10, pp. 5295--5305, 2014.

\bibitem{zhan2015heritability}
L~Zhan et~al.,
\newblock ``Heritability of brain network topology in 853 twins and siblings,''
\newblock in {\em ISBI}, 2015, pp. 449--453.

\bibitem{van2013genetic}
MP~van~den Heuvel et~al.,
\newblock ``Genetic control of functional brain network efficiency in
  children,''
\newblock {\em Eur Neuropsych}, vol. 23, no. 1, pp. 19--23, 2013.

\bibitem{fu2015genetic}
Y~Fu et~al.,
\newblock ``Genetic influences on resting-state functional networks: {A twin
  study},''
\newblock {\em Hum Brain Mapp}, vol. 36, no. 10, pp. 3959--3972, 2015.

\bibitem{yang2016genetic}
Z~Yang et~al.,
\newblock ``Genetic and environmental contributions to functional connectivity
  architecture of the human brain,''
\newblock {\em Cereb Cortex}, vol. 26, no. 5, pp. 2341--2352, 2016.

\bibitem{salvador2005neurophysiological}
R~Salvador et~al.,
\newblock ``Neurophysiological architecture of functional magnetic resonance
  images of human brain,''
\newblock {\em Cereb Cortex}, vol. 15, no. 9, pp. 1332--1342, 2005.

\bibitem{arslan2015joint}
S~Arslan, S~Parisot, and D~Rueckert,
\newblock ``Joint spectral decomposition for the parcellation of the human
  cerebral cortex using resting-state fmri,''
\newblock in {\em IPMI}, 2015, pp. 85--97.

\bibitem{parisot2016grampa}
S~Parisot et~al.,
\newblock ``{GraMPa: Graph-based} multi-modal parcellation of the cortex using
  fusion moves,''
\newblock in {\em MICCAI}, 2016, pp. 148--156.

\bibitem{tymofiyeva2014brain}
O~Tymofiyeva et~al.,
\newblock ``Brain without anatomy: construction and comparison of fully
  network-driven structural mri connectomes,''
\newblock {\em PloS One}, vol. 9, no. 5, 2014.

\bibitem{takerkart2014graph}
S~Takerkart et~al.,
\newblock ``Graph-based inter-subject pattern analysis of {fMRI} data,''
\newblock {\em PloS One}, vol. 9, no. 8, 2014.

\bibitem{glasser2013minimal}
MF~Glasser et~al.,
\newblock ``The minimal preprocessing pipelines for the human connectome
  project,''
\newblock {\em NeuroImage}, vol. 80, pp. 105--124, 2013.

\bibitem{thirion2014fmri}
B~Thirion et~al.,
\newblock ``Which {fMRI} clustering gives good brain parcellations?,''
\newblock {\em Front in Neurosc}, vol. 8, pp. 167, 2014.

\bibitem{rubinov2010complex}
M~Rubinov and O~Sporns,
\newblock ``Complex network measures of brain connectivity: uses and
  interpretations,''
\newblock {\em NeuroImage}, vol. 52, no. 3, pp. 1059--1069, 2010.

\bibitem{riesen2007bipartite}
K~Riesen, M~Neuhaus, and H~Bunke,
\newblock ``Bipartite graph matching for computing the edit distance of
  graphs,''
\newblock in {\em Graph-Based Representations in Pattern Recognition}, pp.
  1--12. Springer, 2007.

\end{thebibliography}

\end{document}